\pdfoutput=1
\documentclass[%
 reprint,
superscriptaddress,
nobibnotes,
 amsmath,amssymb,
 aps,
prb,
floatfix,   
]{revtex4-2}

\usepackage{graphicx}
\usepackage{dcolumn}
\usepackage{bm}
\usepackage{color,soul}
\usepackage[T1]{fontenc}
\usepackage[backref=none, breaklinks=True]{hyperref} 
\usepackage[usenames,dvipsnames]{xcolor}
\usepackage[framemethod=TikZ]{mdframed}
\newcommand\myshade{85}
\colorlet{mylinkcolor}{violet}
\colorlet{mycitecolor}{YellowOrange}
\colorlet{myurlcolor}{Aquamarine}
\hypersetup{
  linkcolor  = mylinkcolor!\myshade!black,
  citecolor  = mycitecolor!\myshade!black,
  urlcolor   = myurlcolor!\myshade!black,
  colorlinks = true,
}

\usepackage{hyperref}
\usepackage[mathlines]{lineno}

\begin{document}

\preprint{APS/123-QED}

\title{A Quantum-Inspired Binary Optimization Algorithm for Representative Selection}

\author{Anna G. Hughes}
\affiliation{%
Agnostiq Inc.,
  325 Front St W, Toronto, ON M5V 2Y1}
\author{Jack S.  Baker}%
\affiliation{%
Agnostiq Inc.,
  325 Front St W, Toronto, ON M5V 2Y1}
\author{ Santosh Kumar Radha}%
\email{research@agnostiq.ai}
\affiliation{%
Agnostiq Inc.,
  325 Front St W, Toronto, ON M5V 2Y1}
\date{\today}

\begin{abstract}
Advancements in quantum computing are fuelling emerging applications across disciplines, including finance, where quantum and quantum-inspired algorithms can now make market predictions, detect fraud, and optimize portfolios. Expanding this toolbox, we propose the selector algorithm: a method for selecting the most representative subset of data from a larger dataset. The selected subset includes data points that simultaneously meet the two requirements of being maximally close to neighboring data points and maximally far from more distant data points where the precise notion of distance is given by any kernel or generalized similarity function. The cost function encoding the above requirements naturally presents itself as a Quadratic Unconstrained Binary Optimization (QUBO) problem, which is well-suited for quantum optimization algorithms - including quantum annealing. While the selector algorithm has applications in multiple areas, it is particularly useful in finance, where it can be used to build a diversified portfolio from a more extensive selection of assets. After experimenting with synthetic datasets, we show two use cases for the selector algorithm with real data: (1) approximately reconstructing the NASDAQ 100 index using a subset of stocks, and (2) diversifying a portfolio of cryptocurrencies. In our analysis of use case (2), we compare the performance of two quantum annealers provided by D-Wave Systems.
\end{abstract}

\maketitle


\section{Introduction}

The task of choosing representative samples from a larger collection of data, often referred to as representative selection, has various advantages over studying the dataset as a whole (e.g., \cite{frey2007, yue2008, rodriguez2014, prasad2014}). Representative selection can reduce the size and complexity of the data, simplifying data analysis and processing and reducing the memory cost of storing data. The computational efficiency of data modeling, such as classifier training and model application, can be significantly improved. Representative selection has been implemented in various subjects ranging from computer vision \cite{elhamifar2012see} and language processing \cite{zhao2017fuzzy} to protein analysis \cite{hobohm1992selection}. A variety of representative selection procedures have been offered to reduce the volume of training data for some specific supervised learning classifiers  \cite{garcia2008memetic,pkekalska2006prototype}. In addition to the methods that require additional knowledge for representative selection, there has been a growing interest in unsupervised approaches to finding representative samples  \cite{garcia2012prototype}.

In this work, we implement an unsupervised representative selection algorithm. An unlabeled dataset may contain some unknown number of classes, or the data in the set could be unclustered or only very loosely clustered into different categories according to some notion of similarity. The algorithm aims to find a subset of representative data points from a larger dataset. The selected subset should include only data points dissimilar to one another and similar to unselected neighboring data points. More concretely, our selector algorithm finds the $k$ least similar points in a sample of $n$ total clustered data points. This is done by both \textit{maximizing} the distance between the chosen data and all other data in the dataset while also \textit{minimizing} the distance between selected data and the other similarly clustered data. The returned $k$ points are then both representative of the data clusters and maximally distant from the other groups. This method can be applied across disciplines, but in this paper, we explore an application in finance: diversification. 

Diversification is a crucial strategy when building a robust portfolio. To mitigate the risk of interdependent components performing poorly in a market downturn, it is critical to invest in assets that are not strongly correlated to one another. There are many portfolio diversification strategies, from a naive $1/N$ rule \cite{demiguel2009} to more complex methods such as portfolio dimensionality or a Bayesian approach  \cite{brochu2010, barkhagen2019}. Many diversification methods are designed to weigh assets to minimize the portfolio's overall variance deliberately. The recent popularity of machine learning in quantitative finance has enabled researchers to define new diversification methods. In this paper, we use our selector algorithm as a method for the diversification of assets. In this framework, a portfolio is well-diversified when each asset is maximally dissimilar to one another and representative of their respective sectors or other similarly performing assets. The core framework of our selector algorithm emerges as quadratic unconstrained binary optimization (QUBO) problem, which can be tackled with a range of metaheuristics \cite{glover1986, glover1999, beasley1999, dunning2018} utilizing classical and quantum computation. Because of the large number of variables considered in diversification (and other selective representation problems), the resulting QUBO objective has many binary variables which, in the quantum setting, translates to the requirement of a large number of qubits. Although gate-model quantum computers are continually scaling up qubit numbers, presently, \textit{quantum inspired} hardware like quantum annealers already meet this requirement (although the qubits are non-universal). Subsequently, we regard our selector algorithm as quantum inspired. 

\section{The Selector Algorithm}

The selector algorithm is designed to pick out unique and representative data points from a larger dataset by finding low-cost solutions to a QUBO objective function. This function is constructed such that low-cost solutions maximize some notion of distance between selected data points and all other data, ensuring that chosen data points are unique while simultaneously minimizing the distance between selected and similarly clustered data. This ensures that each of the selected $k$ data points from a dataset containing $n>k$ points represent data that are similarly clustered with nearby points whilst each of the $k$ data points are \textit{not} similar to each other.

The input data are provided as a matrix $\mathbf{Y}$, where each row is a vector representative of the $i^{\text{th}}$ data point

\begin{equation}
    \mathbf{Y} = \begin{bmatrix}
           \vec{y}_{1} \\
           \vec{y}_{2} \\
           \vdots \\
           \vec{y}_{i} \\
           \vdots \\
           \vec{y}_{n}
         \end{bmatrix} = \begin{bmatrix}
           y_{1}^{(1)} & y_{1}^{(2)} & y_{1}^{(3)} & \dots \\
           y_{2}^{(1)} & y_{2}^{(2)} & y_{3}^{(3)} & \dots \\
            & \vdots \\
            y_{i}^{(1)} & y_{i}^{(2)} & y_{i}^{(3)} & \dots
            \\
            & \vdots \\
           y_{n}^{(1)} & y_{n}^{(2)} & y_{n}^{(3)} & \dots
         \end{bmatrix}.
\end{equation}

where $n$ is the size of the dataset. Although any metric can be used to evaluate the distance between each data point, throughout this paper, the choice of distance metric is the Euclidean distance unless stated otherwise, 

\begin{equation}
    d_{ij}(\vec{y}_i, \vec{y}_j) = \sqrt{\sum_m \left(\vec{y}_{i}^{(m)} - \vec{y}_{j}^{(m)} \right)^2}.
    \label{eq:dist_ij}
\end{equation}

Alternative distance metrics include standardized Euclidean, cosine, Minkowski, etc. The distance $d$ is formally a kernel function, which in principle could also be computed and/or learned using quantum/classical neural networks  (e.g., \cite{radha2022}). It is therefore possible to create a cross-paradigm variation of the selector algorithm where the $d$ is evaluated on a gate model device as proposed in \cite{radha2022}, and the QUBO objective is approximately solved using a quantum annealer. This however is beyond the scope of this work.

The QUBO objective is defined as

\begin{equation}
\label{eqn:cost_function}
    \mathcal{C}(\vec{x}) =  \frac{1}{2k} \vec{x} \mathbf{d} \vec{x}^T - \frac{1}{n} \vec{x} \mathbf{d} \vec{1}^T + A \left( \sum_{i=1}^{n} x_i - k \right)^2,
\end{equation}

where $\vec{1}$ is the $n$-length vector of ones, $\mathbf{d}$ is the $n \times n$ distance matrix with elements $d_{ij}$ as given in Equation \ref{eq:dist_ij}(or another user-defined distance measure), $\vec{x}$ is the $\{0, 1\}^n$ vector of binary variables and $A$ is a penalty scaling factor used to enforce the equality constraint $\sum x_i = k$. The first term on the RHS of Equation \ref{eqn:cost_function} represents the distance between the selected points and other points in the cluster, ensuring that selected points are representative of their cluster, the second term represents the distance between selected points and all others in the data set, and the last term enforces the equality constraint (i.e., a penalty is applied to the cost function if more than $k$ data points are chosen). This optimization problem takes on a QUBO form which are, in general, NP-hard. A more detailed description of QUBO models is given in Appendix \ref{appendix:QUBO} and a formal extension to weighted selection (i.e, where $\vec{x}$ can take on a number discrete values not limited to $\vec{x} \in \{0, 1 \}^n$) is given in Appendix \ref{appendix:weights}. Now in this form, the problem becomes approachable with approximate metaheuristic algorithms, including quantum annealing, used in Section \ref{sec: solvers}, which we discuss qualitatively in Appendix \ref{appendix:annealing}

\begin{figure}
\centering
 \includegraphics[width=\linewidth]{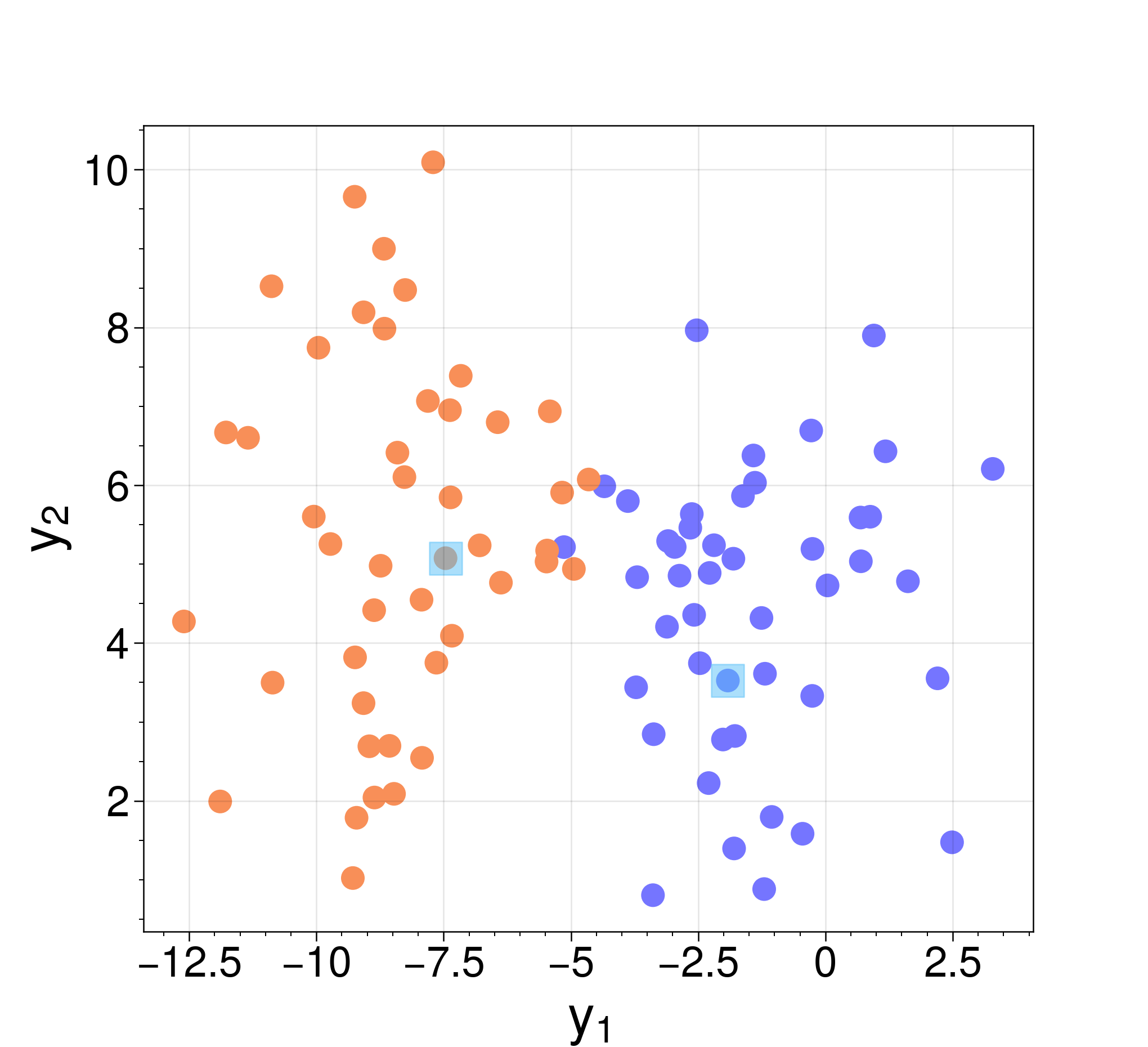}
\caption{An example of the selector algorithm’s performance on clustered data points. The clusters are generated by randomly choosing data points from a Gaussian distribution on a two-dimensional plane centered on two points. Each blob contains 90 data points, with a standard deviation of 2. The selector algorithm was tasked with choosing two representative points from the complete dataset using the \textit{qbsolv} solver. The algorithm-selected points are highlighted with blue squares, indicating that the algorithm successfully chose visually representative points from each cluster.}
\label{fig:blobs}
\end{figure}

\section{Experiments with synthetic data}

To demonstrate the basic functionality of the selector algorithm, we use it to select points from two synthetic datasets. One dataset contains simple and obviously clustered data points (in terms of the Euclidean distance), while the second dataset contains time series data - data points organized in a chronologically ordered sequence. We show that in both cases, the selector algorithm makes reasonable, representative choices. In the first application, we use the selector algorithm to choose 2 data points from a clustered array of points. We generate two clusters, or blobs, by randomly choosing data points from a Gaussian distribution on a two-dimensional plane centered on two points. Each blob contains 90 data points, with a standard deviation of 2. The coordinates of all data points are used as input for the selector algorithm, which then chooses $k$ representative points by minimizing the cost function in Equation \ref{eqn:cost_function} using the D-Wave decomposing solver \textit{qbsolv} \cite{booth2020, qbsolv} which uses a modified tabu algorithm \cite{glover1990} to minimize the objective function. In this example, we have generated two clusters of data points and tasked the selector algorithm with choosing $k=2$ representative points. The blobs and selected points are shown in Figure \ref{fig:blobs}; as expected, the selector algorithm successfully chose one point from each cluster. Since the choice of distance is euclidean, the chosen point closely resembles the approximate center of the cluster. This visual representation might not always be the case for all metrics; for example, metrics like Jensen–Shannon divergence might have no visually discernible centers.

The selector algorithm can similarly choose data points of arbitrarily high dimensions. We demonstrate this in two examples: choosing points from an array of trigonometric curves and choosing from an array of stochastic differential equation (SDE) time series generated randomly. In each case, the data are generated synthetically but clustered into two different classes: sine or cosine in the trigonometric case and time series generated from coupled Brownian process in the SDE case. In both cases, the selector algorithm is tasked with choosing two representative data points; in the trigonometric case, a solution is considered successful if the selector algorithm chooses one sine and one cosine curve, and in the SDE case, a solution is successful if the selector algorithm chooses representatives from the distinct processes.  

\begin{figure*}
\centering
 \includegraphics[width=\linewidth]{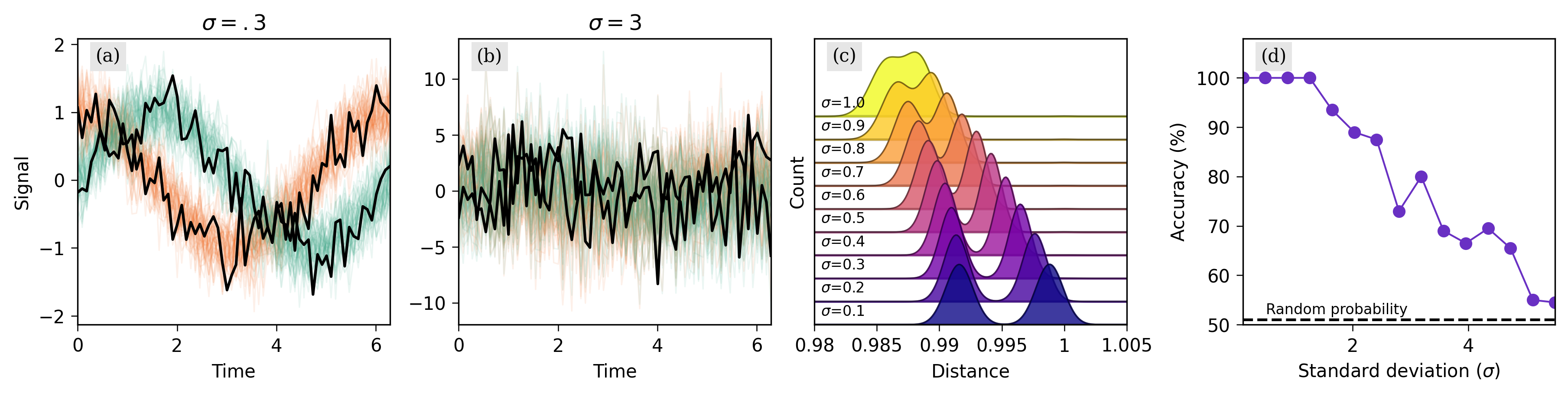}
\caption{Two clusters of 50 sine and cosine curves generated from Equation \ref{eqn:sincos} are used as input for the selector algorithm, with varying values of the standard deviation of the Gaussian noise. \textit{(a)}: 50 sine and cosine time series curves with introduced Gaussian noise of $\sigma = 0.3$ are plotted in green and orange, respectively. The algorithm-selected curves are plotted in black, showing that the algorithm chose one curve from each cluster. \textit{(b)}: 50 sine and cosine time series curves are plotted with introduced Gaussian noise increased to $\sigma = 3.0$. As before, the algorithm-selected curves are plotted in black, but the clusters are no longer distinguishable because the amplitude of the noise is too great. \textit{(c)}: the density of curves are shown as a function of distance, where each horizontal line extending upward represents increasing noise standard deviation. As expected, the two clusters of curves are clearly separable into two distinct peaks at low $\sigma$, but start to merge as $\sigma$ increases. \textit{(d)}: the accuracy of chosen solutions is plotted as a function of $\sigma$ in the bottom right panel. Accuracy here is measured as the percentage of algorithm solutions that contained one sine and one cosine curve out of 200 total trials. At low $\sigma$, the selector algorithm chooses solutions with 100\% accuracy. While that accuracy decreases at higher $\sigma$, even at $\sigma > 3$ where the two sets of curves are indistinguishable, the selector algorithm can pick out curves from each cluster with better-than-random accuracy. All optimizations in this example were performed using the \textit{qbsolv} solver.}
\label{fig:sincos}
\end{figure*}



The trigonometric time series are generated using, 
\begin{equation}
\begin{split}
    y^{\mathrm{sine}}_i = sin(t_i) + n_i \\
    y^{\mathrm{cosine}}_i = cos(t_i) + n_i,
\label{eqn:sincos}
\end{split}
\end{equation}
where $t_i$ ranges from 0 to $2 \pi$ and $n_i$ is artificial random noise pulled from a Gaussian distribution with mean $\mu = 0$ and standard deviation $\sigma$ ranging from $\sigma = 0.1$ to $\sigma = 5.5$. As $\sigma$ increases, there is little difference between the sine and cosine curves because the introduced noise dominates the amplitude of the curves. We generate 50 sine and 50 cosine time series with 10 different $\sigma$ values and task the selector algorithm with choosing $k=2$ representative data points in each case. We expect the selector algorithm to select one sine and one cosine data point when $\sigma$ is low, but as $\sigma$ is increased, it should be harder for the selector algorithm to distinguish the two clusters.

The results are shown in Figure \ref{fig:sincos}. In the first two panels (a,b), clusters of sine (green) and cosine (orange) curves are plotted against time, where $t$ spans from $0$ to $2 \pi$. The curves in the top left panel have $\sigma = 0.3$, while the curves in the top right panel have $\sigma = 3.0$. In each plot, the 2 representative curves chosen by the selector algorithm are plotted in black. As expected, the $\sigma = 0.3$ time series curves are easily distinguishable, with the algorithm choosing one of each. In (b), where the standard deviation is increased to $\sigma = 3$, the sine and cosine curves are no longer visually distinguishable. 

In (c), density plots of the elements of the correlation distance matrix are shown. Each horizontal line extending upward represents increasing noise standard deviation, from $\sigma = 0.1$ to $\sigma = 1.0$. As expected, the two clusters of curves are clearly separable into two distinct peaks at low $\sigma$, but start to merge as $\sigma$ increases. This can measure the distinguishability (or lack thereof) of the clusters in the dataset as a function of $\sigma$. 

In (d), the accuracy of chosen solutions is plotted as a function of $\sigma$. 
Accuracy here is measured as the percentage of algorithm solutions that contained one sine and one cosine curve out of 200 total trials. As expected, at low $\sigma$, the selector algorithm chooses solutions with 100\% accuracy. While that accuracy decreases at higher $\sigma$, even at $\sigma > 3$ where the two sets of density curves (figure (c)) are indistinguishable, the selector algorithm can pick out curves from each cluster with better-than-random accuracy. This demonstrates the robustness of the selector algorithm in choosing representative data points even as the separation between clusters vanishes.

Finally, we test the performance of the selector algorithm at choosing solutions from the solution of a two-dimensional SDE given by 
\begin{equation}
d x_{i}(t)=\mu_{i} x_{i}(t) d t+\sum_{j=1}^{2} \sigma_{i j} x_{i}(t) d W_{j}(t)\label{eq:sde},
\end{equation}
where $\mu_{i}$ and $\sigma_{i j}$ are the drift vector and volatility matrix respectively and $W_{j}(t)$ is the standard Brownian motion. For the case of simulation, we choose $\mu_i$ and $\sigma$ randomly from a uniform distribution $[-1,1]$ and numerically generated multiple two-dimensional time series.

\begin{figure}
\centering
 \includegraphics[width=\linewidth]{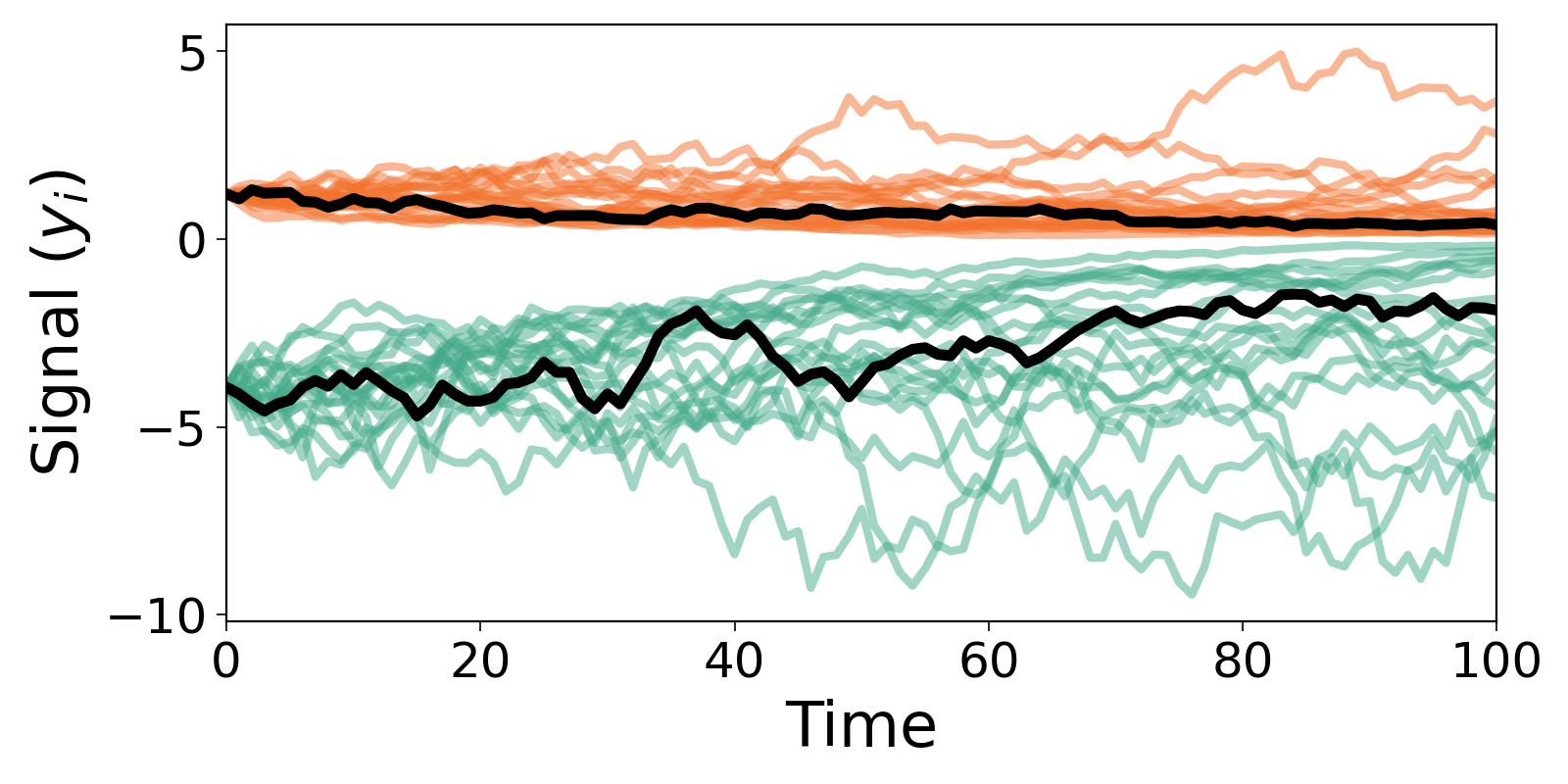}
\caption{Two clusters, each containing 50 synthetic financial time series, generated as given in \autoref{eq:sde}, are plotted in orange (positive) and teal (negative). Each cluster is created with a random correlation chosen from a Gaussian distribution of $\mu = 0$ and $\sigma = 4$. For each set of curves, the volatility and return rates are randomly chosen from a Gaussian distribution with $\mu = -1$ and $\sigma=1$. The selector algorithm was tasked with choosing 2 points from the full dataset. Plotted in black are the two selected curves, showing the selector algorithm’s ability to choose representative time series from synthetic financial data reliably.}
\label{fig:SDE}
\end{figure}

While the trigonometric time series showed us the ability of the selector algorithm to choose representative data points from loosely clustered or unclustered data, in this example, we show the ability of the selector algorithm to evaluate simulated financial data. As before, the selector algorithm was given an array containing all-time series and tasked with choosing (apriori known) $k=2$ representative curves. The two sets of correlated SDE time series are plotted in Figure \ref{fig:SDE}, with the selector algorithm's choices plotted in black. The selector algorithm successfully chose one curve from each time series cluster, demonstrating its ability to evaluate financial time series data such as daily returns.

\section{Use Case: Building a Diversified Portfolio}

Diversifying a portfolio by investing in uncorrelated assets is an approach to mitigate risks associated with downturns in specific markets or unexpected crises. Because the selector algorithm is designed to pick out data points that are both distinct and representative, it is particularly useful for building a diverse portfolio from a more extensive list of assets. Quantum annealers are particularly well-suited to address QUBO problems at this scale. That is, it is possible for the large number of variables to be mapped to the large number ($\sim 1000$) of non-universal qubits available on present generation quantum annealers. In \autoref{subsec:port:sim}, we will start by using the selector algorithm as a portfolio diversifier aiming to approximate the behaviour of the NASDAQ 100 index using a smaller subset of assets. For this task we use a classical large scale QUBO solver which is presently practical to use. We then proceed to \autoref{subsec:port:solver}, where we perform experiments with the algorithm using D-Wave's quantum annealers. That is, we benchmark and compare the performance of two quantum devices for the task of diversifying a cryptocurrency portfolio.

\subsection{Reconstructing the NASDAQ 100 with a classical QUBO solver}\label{subsec:port:sim}

The selector algorithm can build a diverse portfolio by selecting a representative subset of stocks from a market index. In this first application on real financial data, we use the selector algorithm to approximately reconstruct the NASDAQ 100 by choosing a subset ($S_k$) of $k$ stocks from all 102 stocks in the market index. We perform the optimization with the D-Wave decomposing solver \textit{qbsolv} \cite{booth2020, qbsolv}.

To choose stocks from the NASDAQ 100 index, we treat the daily returns from each stock as individual data points in our array. Vectors $\vec{y}_i$ are composed of the daily returns from each stock in the index over one trading year, or 253 days, starting on 2021-02-01 (YYYY-MM-DD) and ending on 2022-02-01 (YYYY-MM-DD). This results in a $102 \times 253$ matrix \textbf{Y},
\begin{equation}
    \mathbf{Y} = \begin{bmatrix}
           \overrightarrow{\vec{y}_1} \\
           \overrightarrow{\vec{y}_2} \\
           \overrightarrow{\vec{y}_3} \\
           \vdots \\
           \overrightarrow{\vec{y}_n}
         \end{bmatrix} = \begin{bmatrix}
           \overrightarrow{AAPL} \\
           \overrightarrow{ABNB} \\
           \overrightarrow{ADBE} \\
           \vdots \\
           \overrightarrow{ZS}
         \end{bmatrix}
\end{equation}

In this familiar form, the data are ready for evaluation in the cost function (Equation \ref{eqn:cost_function}), where $n = 102$ is the total number of vectors in matrix \textbf{Y}, $k$ is the number of desired stocks selected by the algorithm, and the elements of $\mathbf{d}$ are given by the correlation distance, 
\begin{equation}
    d_{ab} = 1 - \frac{\sum(a_i - \bar{a}) \cdot \sum(b_i - \bar{b})}{\sqrt{\sum (a_i - \bar{a})^2 \sum (b_i - \bar{b})^2}},
\end{equation}
where $\bar{a}$ and $\bar{b}$ are the mean values of vectors $\vec{a}$ and $\vec{b}$.

The chosen stocks can come arbitrarily close to a complete reproduction of the market index; as the number of selected stocks $k$ increases, the combined data from the chosen stocks comes closer and closer to an exact reproduction of the index itself. However, the goal is only approximately to reproduce the index with a smaller number of stocks, so choosing a large number is not necessarily advantageous and is a hyperparameter for the end-user.



In all following analyses, we use only the binary case, where the weighted subspace is restricted to only binary values, $\vec{w} \rightarrow \vec{x} \in \{0, 1\}^n$ (see discussion in Appendix \ref{appendix:weights} for the general discrete weighted case). Like in the previous examples, this means that specific data points are either chosen or not chosen, with no inbetween. This is an NP-hard problem \cite{barahona1982}. To evaluate the performance of the binary selector algorithm in reproducing the NASDAQ 100, we create a proxy NASDAQ 100 index by linearly combining all stocks and averaging the resulting vector. Note that while the NASDAQ 100 index is market value-weighted, for simplicity, the stocks in our NASDAQ 100 index are equally weighted.

\begin{figure*}
\centering
\includegraphics[width=\linewidth]{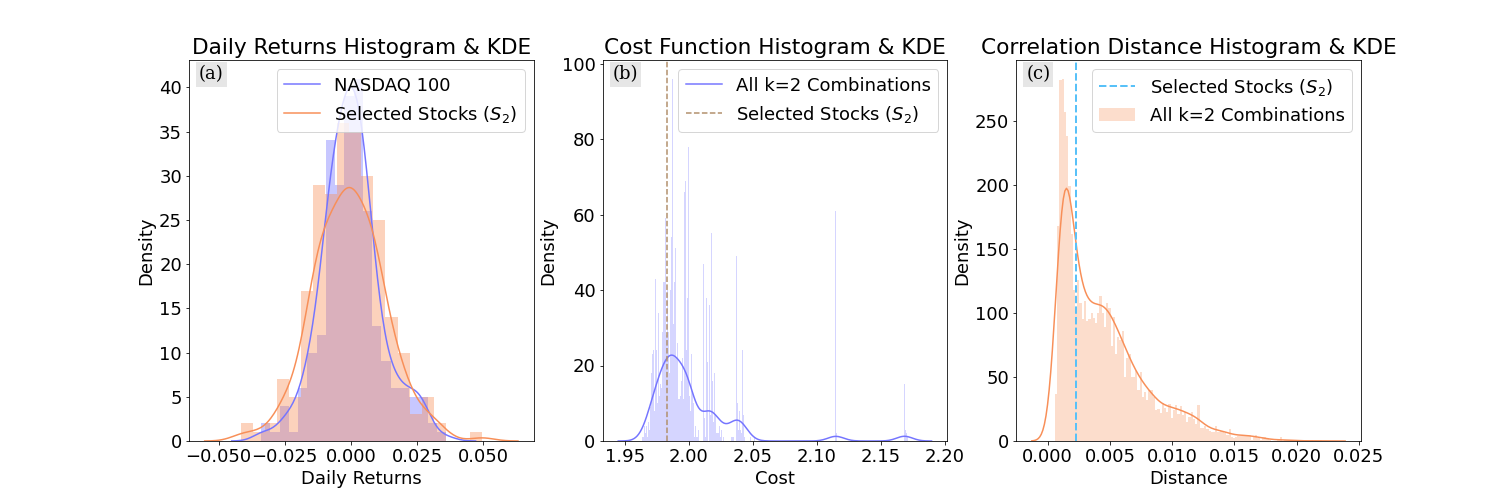}
\caption{\textit{(a)} Histograms and corresponding kernel density estimate (KDE) plots of the daily returns from the proxy NASDAQ 100 index and the combined index from $k=2$ stocks ($S_2$) selected by the algorithm using the \textit{qbsolv} solver. \textit{(b)} Histogram and corresponding KDE plot showing the value of the cost function for all combinations of 2 stocks in the NASDAQ 100 index, where the dashed vertical line shows the value of the cost function for the algorithm-selected $k=2$ stocks. \textit{(c)} Histogram and corresponding KDE plot showing the correlation distance for all combinations of 2 stocks in the NASDAQ 100 index, where the dashed vertical line shows the distance for the algorithm-selected $k=2$ stocks. }
\label{fig:kdes}
\end{figure*}

We initially use the selector algorithm to choose two stocks from our input matrix \textbf{Y} of returns. The algorithm-chosen stocks are \texttt{Fox Corporation (`FOX’)} and \texttt{Synopsys, Inc. (`SNPS’)}. We use these stocks to create $S_{k=2}:= S_2$, a data point composed of the averaged linear combination of our chosen stocks' daily returns.  A histogram showing the average percentage daily returns for both $S_2$ and the proxy NASDAQ 100 index is shown on the left panel of Figure \ref{fig:kdes}. One can see that the stock return profile of the selected $S_2$ stocks is a good surrogate for the NASDAQ 100 already at $k=2$. To further assess the quality of the algorithm's selection, we calculate the value of the cost function and the correlation distance between the proxy NASDAQ 100 index and some data point $S_2$ for all possible combinations of 2 stocks. A histogram showing the range of cost function and correlation distance values are shown in the middle and right panels of Figure \ref{fig:kdes}, respectively. The $S_2$ index had a lower cost function value than $74 \%$ of all cost function values and a correlation distance less than $89 \%$ of all values.

\begin{figure*}
\centering
 \includegraphics[width=0.95\linewidth]{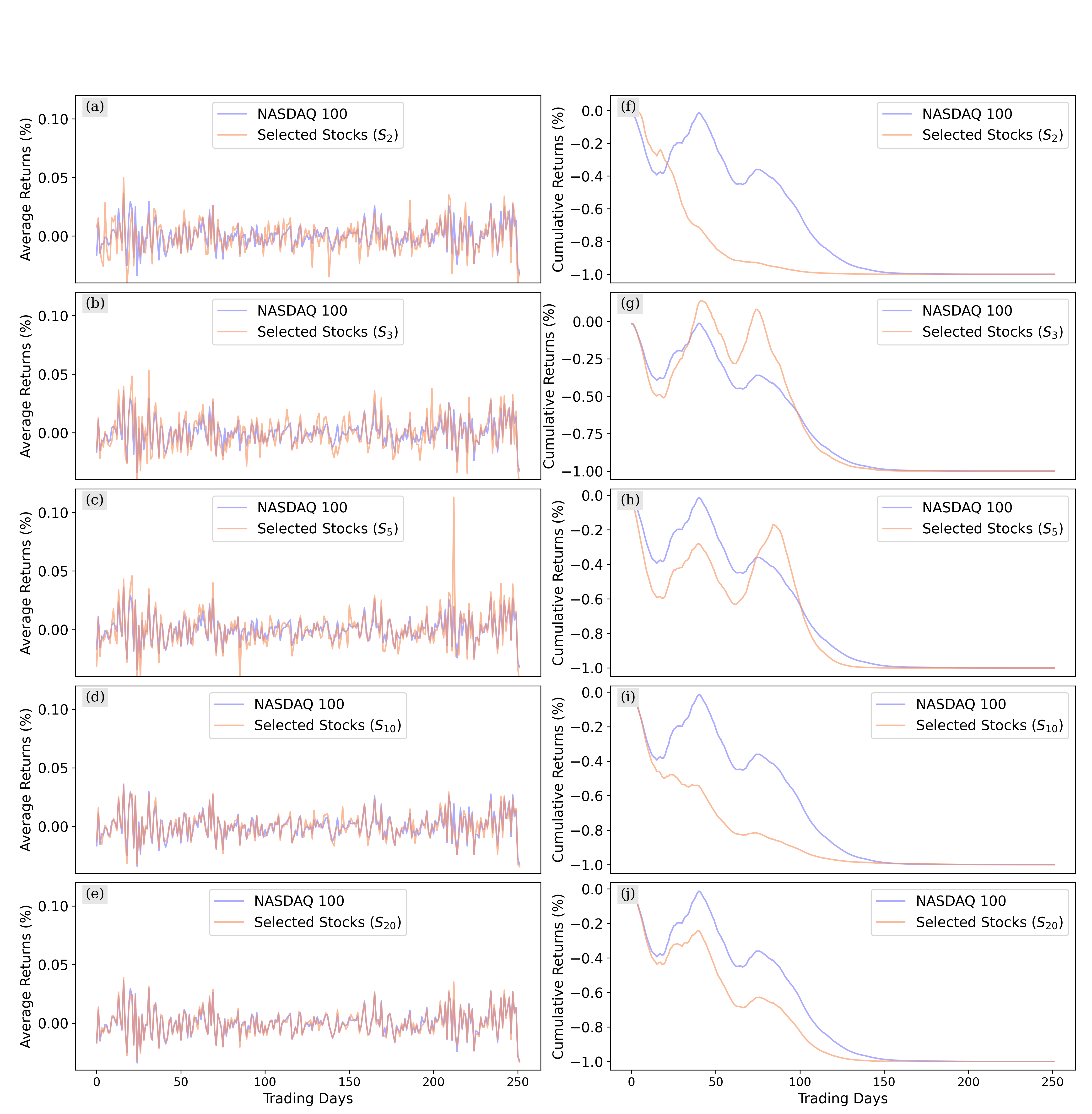}
\caption{\textit{(a)-(e)}: All daily returns from the proxy NASDAQ 100 index (violet) and the index generated from the selector algorithm’s choice of stocks only. The proxy NASDAQ 100 index is created by equally weighting all stocks in the NASDAQ 100 over 2021-02-01 to 2022-02-01 (YYYY-MM-DD), while the selector algorithm index is created by equally weighting only the stocks chosen by the algorithm. The number of selected stocks $k$ varies, starting from $k=2$ in the first row and increasing to $k=20$ in the final row. As $k$ increases, the selector-created index comes closer to the proxy NASDAQ 100 index, as expected. \textit{(f)-(j)}: The cumulative returns from both the proxy NASDAQ 100 index and the selector algorithm-generated index, where each row represents an increasing number of selected stocks $k$. As the number of selected stocks is increased, the selector algorithm-generated index more closely replicates the features of the proxy NASDAQ 100 index.
}
\label{fig:time_series_rets}
\end{figure*}

In Figure \ref{fig:time_series_rets}, we show the average (left) and cumulative (right) returns from the proxy NASDAQ 100 index plotted with the $S_k$ time series, with each row showing incremental increases in $k$. As expected, when the number of chosen stocks ($k$) increases, the $S_k$ vector comes closer to the proxy NASDAQ 100 index.

\begin{figure}
 \includegraphics[width=7cm]{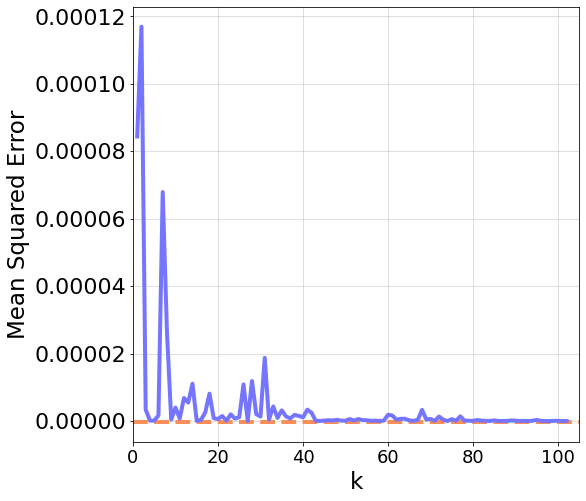}
\caption{The mean-squared error (MSE) between the proxy NASDAQ 100 index and the selector algorithm-generated index as a function of the number of selected stocks k. The proxy NASDAQ 100 index is  created by equally weighting all stocks in the NASDAQ 100 over 2021-02-01 to 2022-02-01 (YYYY-MM-DD). In contrast, the selector algorithm index is created by equally weighting only the stocks chosen by the algorithm. As $k$ increases from 1 representative stock to all 102 stocks in the index, the MSE sharply decreases to zero.}
\label{fig:MSE}
\end{figure}

Finally, we compare the algorithm-selected indices $S_k$ to the NASDAQ 100 index by computing the mean squared error (MSE) between the two. The number of stocks selected $k$ increases steadily until all 102 stocks in the index are included. The MSE as a function of $k$ is plotted in Figure \ref{fig:MSE}, where one can see that NASDAQ 100 can effectively be reproduced by roughly 40 stocks.

\subsection{Diversifying cryptocurrency portfolios with quantum annealers \label{sec: solvers}} \label{subsec:port:solver}

Selecting a diverse portfolio from a limited index can be challenging, especially when the assets cannot be easily divisible into traditional sectors, as in the case of cryptocurrencies (crypto). Unexpected correlations between coins can leave a crypto portfolio vulnerable to the volatility of one asset. The selector algorithm is thus particularly useful for building a diverse portfolio of cryptocurrencies. In this use case, we use the selector algorithm to build a diverse portfolio from the Crescent Crypto Market Index (CCMIX) using quantum annealers. We compare the performance of each device, investigating how well each device satisfied the problem constraints and the quality of solutions.

The two quantum annealing devices used in the following experiments are the D-Wave 2000Q and the D-Wave Advantage. The D-Wave 2000Q, first introduced in 2017, is a 2048-qubit quantum annealing device. The 2000+ qubits are arranged in a Chimera topology with 6016 couplers. The 2000Q system has been used to solve problems ranging from e-commerce listing order to cryptography  \cite{nishimura2019, mengoni2020}. The D-Wave Advantage quantum processing unit (QPU) was released in September 2020 as an updated, more advanced system. It contains 5000+ qubits with 35,000+ couplers - an increase from 6 to 15 couplers per qubit. This increase in qubits and couplers enables the D-Wave Advantage QPU not only to solve larger problems than the D-Wave 2000Q device but also allows for problems with more challenging connectivity to be more easily mappable to the device qubits when compared to the 2000Q Device. The D-Wave Advantage QPU has been used to solve problems such as railway dispatching and molecular unfolding  \cite{domino2021, mato2021}.



The input data to the selector algorithm consists of the daily returns from each coin in the CCMIX over seven months, beginning on 2021-04-01 (YYYY-MM-DD) and ending on 2021-11-11 (YYYY-MM-DD). Depending on the experiment, the selector algorithm chooses some desired number of coins in the array, using the specified quantum annealer to minimize the cost function.

\subsubsection{Constraint Satisfaction with Default Parameters}

\begin{figure}
\centering
 \includegraphics[width=\linewidth]{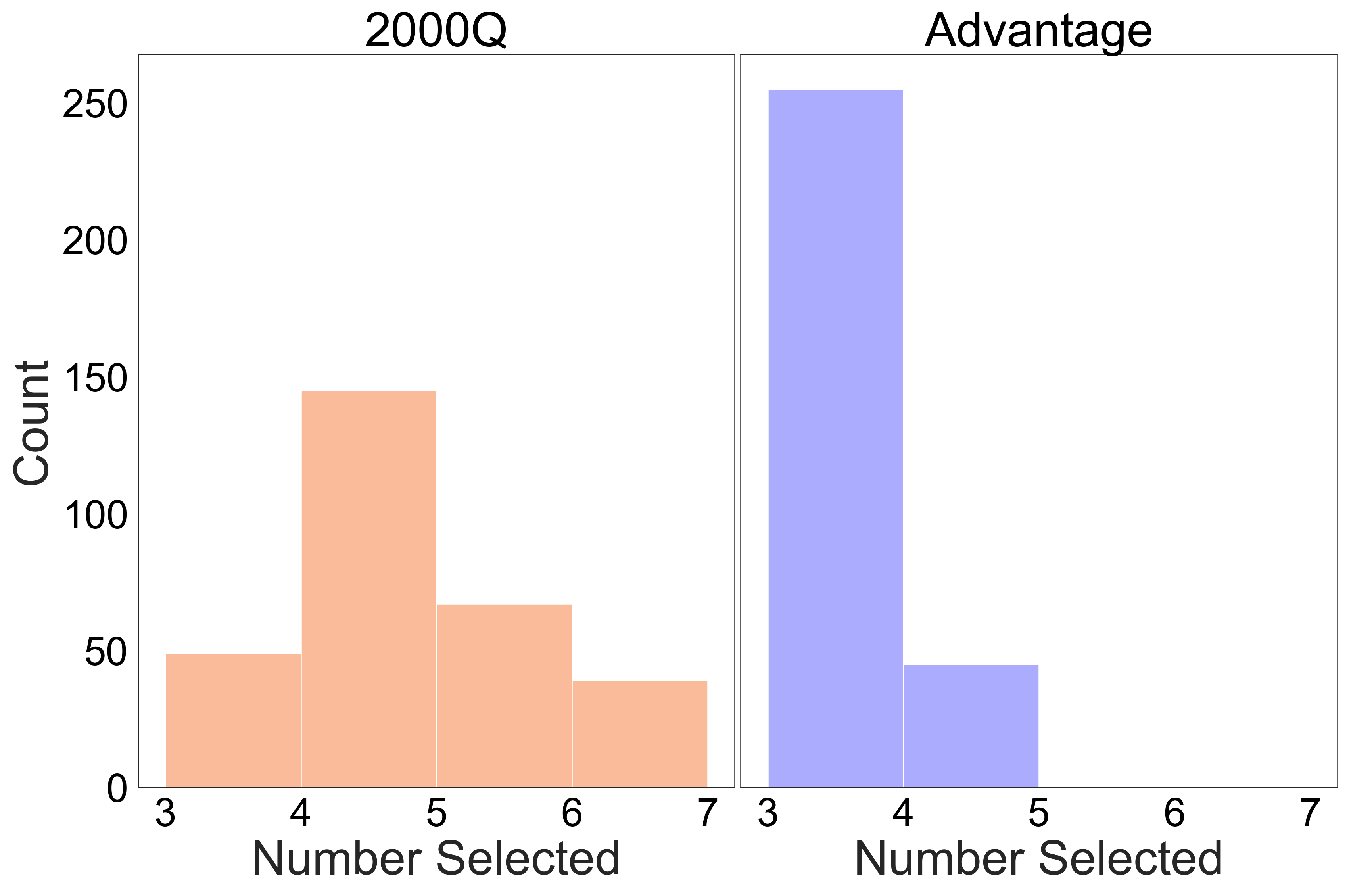}
\caption{The selector algorithm was tasked with choosing three cryptocurrencies from the CCMIX, which contains 18 total options. Optimization was performed using the D-Wave 2000Q device (left) and the D-Wave Advantage device (right), with 300 trials on each QPU. The solutions chosen by the D-Wave 2000Q device satisfied the problem constraint of $k=3$ in only 16\% of its trials, while the D-Wave Advantage device chose solutions that satisfied the problem constraint in $>85\%$ of its trials.}
\label{fig:test1}
\end{figure}

In the first round of experiments, we investigate the ability of each device to choose solutions that satisfy the equality constraint imposed by the penalty term in Equation \ref{eqn:cost_function}. While satisfying this constraint may be trivial for some classical methods, quantum annealers are more limited, especially when the constraints are imposed as penalties \cite{hen2016}. For experiments in this section, we use the default parameters of the annealers, notably leading to an annealing time of $t=20 \, \mu s$. We also use a penalty scaling factor of $A=2$. We run 300 trials on each device, where the selector algorithm was tasked with choosing $k=3$ cryptocurrencies from the 18 coins included in the CCMIX. In all experiments, the number of shots per run is fixed at 1000.

The results are shown in Figure \ref{fig:test1}. There is a significant difference in how well each device satisfies the $k = 3$ constraint. The D-Wave 2000Q device only chose three cryptocurrencies in 49 of the 300 trials -- a success rate of just 16\%. Conversely, the D-Wave Advantage QPU chose three coins in 256 of the 300 runs -- a success rate of over 85\%. There is clearly a marked improvement in performance for the newer D-Wave Advantage QPU even using the default parameters.

\subsubsection{Consistency of Solutions with Changing Annealing Time}

In subsequent experiments, we investigated the impact of the value of the penalty scaling factor $A$ (see Equation \ref{eqn:cost_function}) and the annealing time on constraint satisfaction. We first investigated tuning the penalty scaling factor in 11 runs with 50 repeats for each value of $A$, where the value ranged from $A=0$ to $A=10$ in fixed increments while the annealing time was fixed at $20 \, \mu s$. In these experiments, we saw very little impact on either the quality of the solution (i.e, how low the cost is) or satisfaction of the problem constraint. For the rest of this section, were therefore continue with the default of $A=2$. Finally, we performed 19 trials on each device using annealing times ranging from the default 20 $\mu s$ to 990 $\mu s$, with 50 repeats at each annealing time.

\begin{figure}
\centering
 \includegraphics[width=0.9\linewidth]{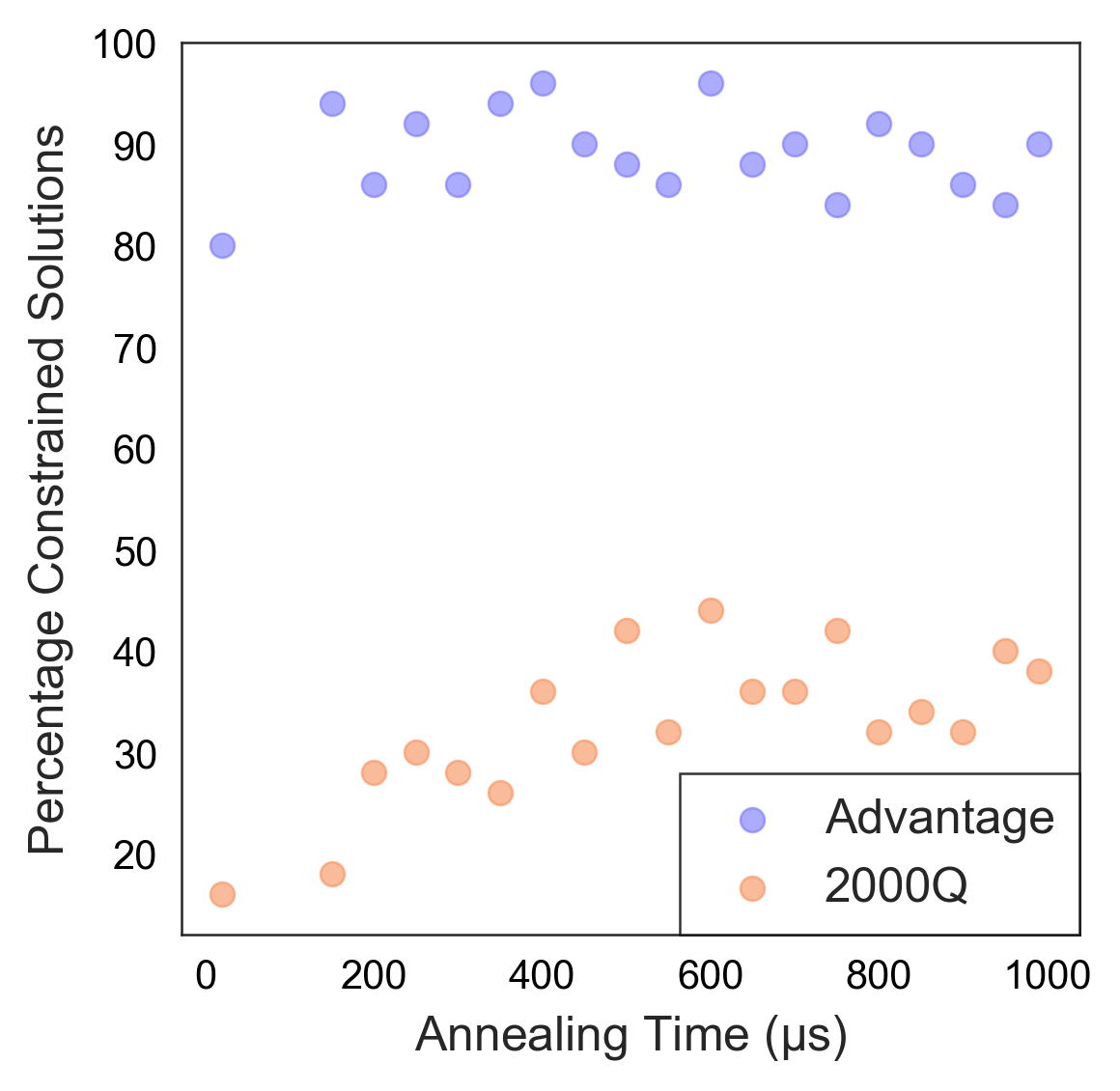}
\caption{We investigate the ability of each device to choose solutions matching problem constraints as a function of annealing time, where the annealing time is increased from $20 \mu s$ to $990 \mu s$. The percentage of solutions matching problem constraints (out of 50 trials at each annealing time) is plotted for the D-Wave 2000Q (beige) and the D-Wave Advantage (violet). The change in annealing time has a negligible effect on the percentage of constraint-satisfying solutions for the Advantage QPU, while higher annealing times seem to enable the 2000Q QPU to satisfy the problem constraint better. However, even at high annealing times, most solutions chosen by the 2000Q device do not satisfy the problem constraint, while $>85\%$ of solutions chosen by the Advantage QPU satisfy the constraint at every annealing time.}
\label{fig:test2}
\end{figure}

As can be seen in Figure \ref{fig:test2}, the value of the annealing time has a negligible effect on the percentage of solutions matching the $k=3$ constraint for the Advantage QPU, which tends towards solutions that satisfy this constraint. Conversely, higher annealing times enable the 2000Q QPU to find more solutions that satisfy the soft constraint. Even at high annealing times, most of the solutions chosen by the 2000Q device do not satisfy the problem constraint, while the Advantage QPU satisfies the constraint $>85\%$ of the time at every annealing time.

The possible impact of the annealing time on the overall value of the cost function ($k=3$) is more difficult to discern. At the intermediate annealing time of 600 $\mu s$ and at the higher annealing time of 900 $\mu s$, both the value of the cost function and the standard deviation between trials hits a low value for the 2000Q QPU. Very little such variation is discernible for the Advantage device.

\subsubsection{Comparing solution quality}

We examine the quality of solutions selected by each device given the constraint that $k = 3$. To understand how the QPU-chosen solutions compare with all possible solutions, we first compute the value of the cost function for every combination of coins. The CCMIX index contains 18 cryptocurrencies, leading to $2^{18} = 262144$ total possible combinations, ranging from choosing 0 coins to the possibility that all 18 coins are chosen. The cost function values are calculated assuming the $k = 3$ constraint, so all combinations containing more or less than three coins are penalized in the cost function. $A=2$ in these experiments. The mean value of the cost function, considering all combinations, is 81.

\begin{figure}
 \includegraphics[width=\linewidth]{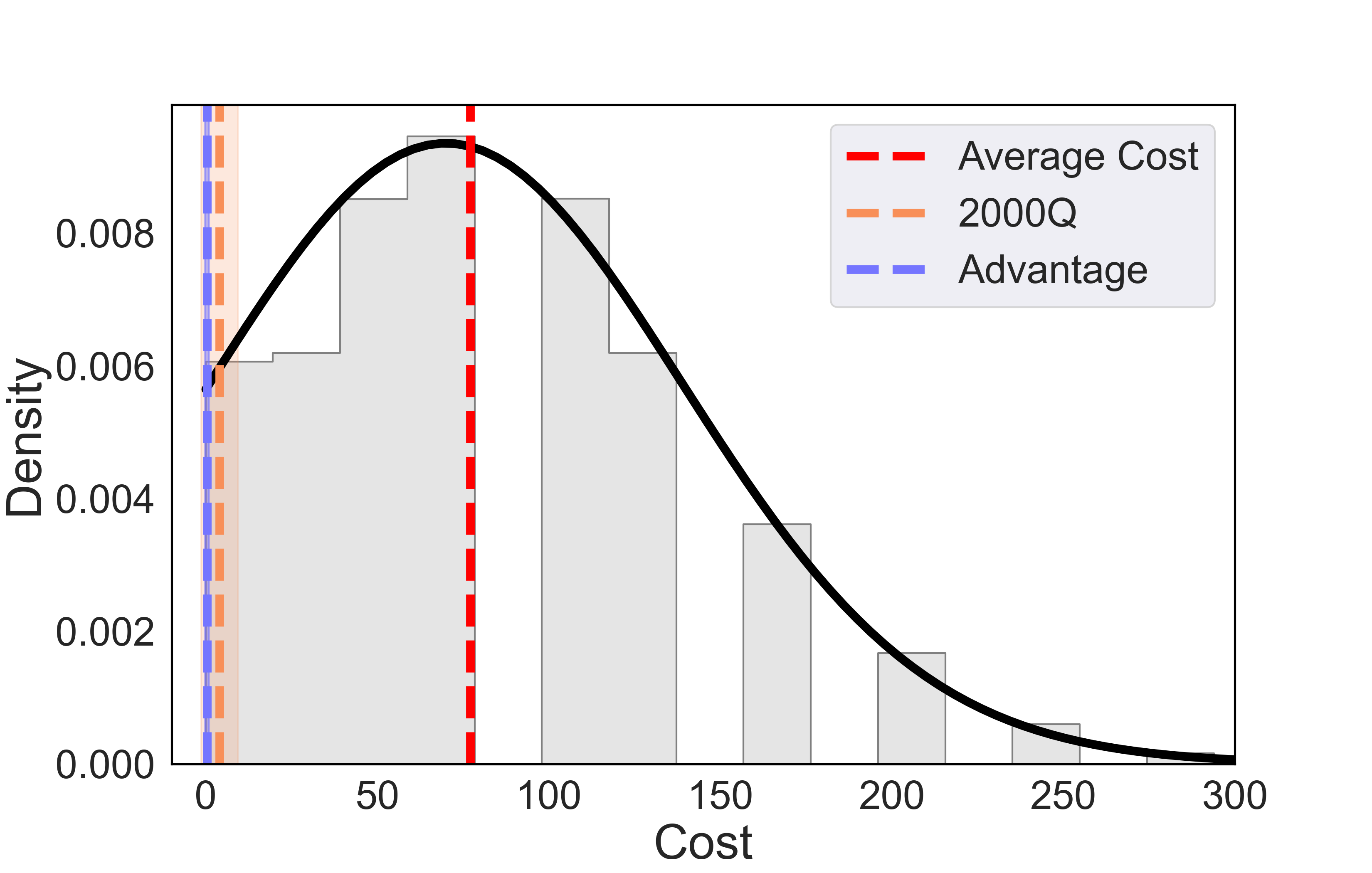}
 \caption{The value of the cost function for all combinations of coins is plotted as a histogram, with the black curve fitted to the distribution. The average value of the cost function, considering all possible combinations, is indicated by the vertical dotted red line. The average cost function value of solutions chosen by the D-Wave 2000Q and Advantage QPUs are indicated by the blue and orange horizontal dotted lines, respectively.}
\label{fig:combos_curve}
\end{figure}

To gauge the performance of quantum devices relative to a randomized solution, we calculate the average value of the cost function for solutions chosen by each device for all trials at each device's optimal annealing time (550 $\mu s$ for 2000Q and $900 \mu s$ for Advantage). For the 2000Q device, the average value of the cost function was 4.02, while the average cost function value for the Advantage QPU was 0.32. The results are shown in Figure \ref{fig:combos_curve}, where a histogram showing the density of cost function values is plotted, with vertical lines indicating the total average cost value and the cost values of the 2000Q and Advantage QPUs.

The average cost function value of both devices was significantly lower than the average value considering all possible solutions. The average value of the cost function for the 2000Q device is in the lowest $4\%$ of possible values, while the average value of the cost function for the Advantage QPU is in the lowest $0.03\%$ of all solutions. Overall, both devices distinctly demonstrate the ability to choose solutions with low values of the cost function. However, comparing devices, the Advantage QPU shows a marked improvement in the ability to choose solutions that match the problem constraints and solutions with lower cost function values. 

\section{Conclusion}

In this work, we devised a selector algorithm for the representative selection of data which relies on solving a QUBO problem. Our selector algorithm chooses data points by maximizing the distance between selected points and all other data points in the dataset and minimizing the distance between selected points and similarly clustered data points. Given the potentially large number of variables in the resulting QUBO problem, the optimization stage of selector algorithm is particularly suited towards large scale QUBO solvers like quantum annealers.

After experimentation with synthetic datasets, in our first practical use case, we used the selector algorithm to approximate the NASDAQ 100 with a subset of stocks in the index. We created a proxy NASDAQ 100 index by linearly combining and averaging all 102 stocks. The selector algorithm chose $k$ stocks from the index, whose performances were combined, averaged, and compared to the proxy 100 indexes. As the number of selected stocks $k$ increased, the algorithm-constructed index more closely resembled the proxy NASDAQ 100 index. We compared the $k=2$ point, $S_2$, to all possible combinations of 2 stocks. We found that the algorithm-selected point $S_2$ had a cost function value in the bottom 26\% of all $k=2$ solutions.

In our second application, we used the selector algorithm to build a diversified portfolio of cryptocurrency assets from the Crescent Crypto Market Index. The optimization was performed using 2 quantum annealers: the D-Wave 2000Q and the D-Wave Advantage. We compared the $k=3$ selection of both devices by evaluating the cost function and comparing that to the value of the cost function of all possible coin combinations in the index. We found that the average cost function value of solutions chosen by the 2000Q device was in the lowest $4 \%$ of all cost function values, while the D-Wave Advantage device's average cost function value is in the lowest $0.03 \%$. The D-Wave Advantage device also more consistently chose solutions matching the problem constraint - with $> 85 \%$ of solutions matching the constraint, while only $16 \%$ of 2000Q solutions matched the constraint. Overall, we saw clear improvement between the newer Advantage QPU and the earlier 2000Q QPU, providing meaningful solutions to the combinatorial optimization problem. 

\bibliography{ms}

\appendix
\section{Quadratic Unconstrained Binary Optimization \label{appendix:QUBO}}

Quadratic Unconstrained Binary Optimization (QUBO) is the optimization of an objective function with binary variables with at-most quadratic order coupling between variables. This approach is powerful and widespread because many combinatorial problems can be written in QUBO form (see Equation \ref{eq:QUBO}) and solved using general purpose QUBO solvers. This is in contrast to other more traditional methods which are tailored to the exact problem at hand. In general, to approximately solve a QUBO problem (the exact problem is NP-hard), we must select an approximately optimal solution from all possible solutions. 

The basic cost or objective function $\mathcal{C}(\vec{x})$ of QUBO is, 
\begin{equation}
    \mathcal{C}(\vec{x}) = \vec{x}^T \mathbf{Q} \vec{x},
    \label{eq:QUBO}
\end{equation}
where $\vec{x} = \{0, 1\}^n$, and \textbf{Q} is an $n \times n$ matrix of constants. The goal of QUBO is to find the arrangement of binary values in $\vec{x}$ that minimizes (or maximizes) the value of $\mathcal{C}(\vec{x})$. Other than the requirement that the values in the vector $\vec{x}$ be binary, there are no formal constraints in this simple QUBO model. However, so-called \textit{soft constraints} can be encoded into the QUBO problem using quadratic penalty terms (for example, see the last term of the RHS of Equation \ref{eqn:cost_function}). Rather than impose rigid constraints on the problem, soft constraints work by discouraging the optimizer from solutions that will incur penalties.

QUBO problems have garnered recent attention due to how easily they are adapted mappable to quantum annealers and gate-model quantum computers.

\section{Quantum Annealing \label{appendix:annealing}}

In metallurgy, annealing strategically applies heat to a material to make it more malleable. Annealing works by initially raising the temperature of a metal above its recrystallization temperature, which causes the atoms in the metal to move around rapidly. The temperature is then slowly lowered, and atoms settle into a new crystalline configuration with the desired properties. Simulated annealing is a machine-learning technique that takes its name from this metallurgic process. Similar to how metallurgic annealing causes atoms to explore new positions rapidly, simulated annealing rapidly samples the solution space of a particular cost function and adopts new configurations as desired. The sampling rate slowly decreases, allowing the annealer to settle into an optimized solution.

In simulated annealing, the solution starts in some initial configuration C and a maximum initial `temperature' of $T_{max}$, where the temperature is a hyperparameter that dictates how randomly the annealer samples the cost function. The state C is then advanced into a new configuration N. The change in the value of the cost function can then be calculated. Depending on the current temperature and the change in the value of the cost function, the new configuration N is either accepted as the new state or rejected. A `good' new configuration N would have a lower cost function value than the previous configuration, indicating that the state has advanced closer to the minimum value. However, even `bad' configurations, where the change in cost function value is positive, can be accepted if the temperature is very high. Just like with real metallurgic annealing, the temperature is then gradually decreased. This means that initially, fewer new configurations are accepted unless the cost function value is significantly lower. This method enables the simulated annealer to sample the cost landscape very broadly and ideally settle on the global minimum rather than any local minimum.

Quantum annealing uses a similar technique but explores quantum states for solution space using the strength of quantum fluctuations as a sampling rate hyperparameter. A quantum annealing cost function is a Hamiltonian, expressing the total energy of the system for each quantum state. Minimizing the cost function means finding the quantum state with the lowest energy. Quantum annealing has the additional benefit of quantum tunneling, where the algorithm can tunnel between different low-energy quantum states, avoiding the need to sample the space in between these states and preventing the algorithm from getting trapped in local minima. As the sampling rate slowly decreases, so is the quantum tunneling radius. Initially, the algorithm is given a wide radius to explore many low-energy quantum states. As the system evolves and the sampling rate decreases, the tunneling radius slowly decreases as the algorithm settles on a final state. Quantum annealing provides an approximate solution, not necessarily an absolute one.

\section{The selector algorithm with weights \label{appendix:weights}}

In the main article, we explore the case where $k$ out of $n$ data points are either ``selected" or ``not selected". This problem is innately QUBO since the weighted contribution $w_i$ to the objective function is restricted to $w_i = x_i \in \{0, 1\}$. We now show how this formalism can be extended to a more general case where $k$ total points are selected, and each selected point is assigned a weight with resolution controlled using the integer discretization number $n_D$. \par

The weight of the $i^{\text{th}}$ data point can be expanded with $n_D$ binary variables,

\begin{equation}
    w_i = w_{\text{min}} + (w_{\text{max}} - w_{\text{min}}) \sum_{j = 1}^{n_D} \frac{2^{j - 1}}{2^{n_D} - 1} X_{ij},
    \label{eq:binary_expansion}
\end{equation}

for minimum/maximum weights $w_{\text{min}/\text{max}}$, respectively. In the context of the selector algorithm, a natural choice is $w_{min} = 0$ and $w_{max} = 1$. $X_{ij} \in \{0, 1\}$ are the elements of the rectangular $n \times n_D$ bit matrix

\begin{equation}
    \mathbf{X} = 
       \begin{bmatrix} 
            X_{1, 1} & \dots & X_{1, n_D} \\
            \vdots & \ddots & \\
            X_{n, 1} &        & X_{n, n_D} 
        \end{bmatrix}.
    \label{eq:bit_matrix}
\end{equation}

Increasing the number of terms used in the expansion of Equation \ref{eq:binary_expansion} (i.e, increasing $n_D$) increases the resolution of $w_i$ which, in the quantum setting, comes at the price of increasing the qubit requirement $N$ as given by $N = n \times n_D$. At $n_D = 1$, the original ``selected" or ``not selected" case is obtained $w_i = X_{1, 1} := x_i$. \par

We now define a new variable $\chi_i$ as the product of the weight and the binary variable $x_i$: $\chi_i = w_i x_i$. Now, within Equation \ref{eqn:cost_function}, for the first two terms on the RHS, we make the change of variable $\vec{x} \rightarrow \vec{\chi}$ and insert a new penalty term on the sum of the elements of $\vec{\chi}$ to yield the objective
\begin{equation}
\begin{split}
    & \mathcal{C}(\vec{\chi}, \vec{x}) =  \frac{1}{2k} \vec{\chi} \mathbf{d} \vec{\chi}^T - \frac{1}{n} \vec{\chi} \mathbf{d} \vec{1}^T +  A \left( \sum_{i=1}^{n} x_i - k \right)^2  \\
    & + B \left( \sum_{i=1}^{n} \chi_i^{} - W \right)^2
\end{split}
\label{eq:quartic_cost}
\end{equation}
where W is the budget constraint on $w_i$: $\sum_i w_i = W$ and $B$ (like $A$) is a penalty scaling factor. As presented, Equation \ref{eq:quartic_cost} is quartic order in binary variables but, following the discussion in \cite{glover2019tutorial}, can always be transformed to quadratic and thus reducible to a QUBO problem. While such a transformation can be formally derived, we use the package \texttt{PyQUBO} \cite{zaman2021pyqubo} to compile the problem to a QUBO form intelligently. The compiled problem can then be solved using metaheuristic optimization algorithms, including quantum annealing. \par

It should also be noted that Equation \ref{eq:quartic_cost} is reducible to Equation \ref{eqn:cost_function} in the main text through some specific choices of the variable. Setting $w_{min} = 0$, $w_{max} = 1$ and $n_D = 1$, each $\chi_i$ reduces to $\chi_i = x_i X_{1, 1} = x_i^2 = x_i$. Now, setting $W = k$, the two quadratic penalty terms can be collected into one and the ``selected" or ``not selected" form of Equation \ref{eqn:cost_function} is obtained.

\end{document}